\documentclass{article}
\usepackage[T1]{fontenc}
\usepackage[utf8]{inputenc}
\usepackage{ismir}
\usepackage{amsmath,url}
\usepackage{graphicx}
\usepackage{booktabs}
\usepackage{multirow}
\usepackage{tikz}
\usepackage{xcolor}
\usepackage{dblfloatfix}
\usepackage{placeins}
\usetikzlibrary{positioning, arrows.meta, fit, calc, decorations.pathreplacing}

\title{Multi-Task Multi-Frame Visual Piano Transcription}

\multauthor
  {Yonghyun Kim$^{\flat,*}$ \hspace{1cm} Hoyeol Sohn$^{\sharp,*}$ \hspace{1cm} Juhan Nam$^\sharp$ \hspace{1cm} Alexander Lerch$^\flat$}
  {$^\flat$ Music Informatics Group, Georgia Institute of Technology, USA\\
  $^\sharp$ Graduate School of Culture Technology, KAIST, South Korea\\
  {\small \{yonghyun.kim, alexander.lerch\}@gatech.edu,
  \{hoyso48, juhan.nam\}@kaist.ac.kr}\\
  \vspace{0.2cm}
  {\small $^*$These authors contributed equally to this work.}
  }
\def\authorname{Y. Kim, H. Sohn, J. Nam, and A. Lerch}

\usepackage{microtype}
\usepackage{paralist}
\usepackage[bookmarks=false,pdfauthor={\authorname},pdfsubject={\pdfsubject},hidelinks]{hyperref}

\begin{document}

\maketitle
\begin{abstract}
Audio-based piano transcription performs well on onset, pitch, and velocity, but the sustain pedal lets sound persist long after key release, so audio systems predict pedal-extended offsets rather than physical key release.
Yet existing Visual Piano Transcription (VPT) systems focus on onset detection from short video windows, offset accuracy lags onset by a wide margin, and note-level velocity has not been reported.
To address these gaps, we present \textbf{V2N (Video to Notes)}, the first complete VPT system: a shared temporal backbone feeds task-specific heads for onset, offset, key hold, and velocity, jointly trained with per-frame supervision rather than only at the window center.  Ablations show that multi-task supervision enables offset and velocity prediction while improving onset accuracy; longer temporal context yields further improvements. V2N sets new state-of-the-art results on PianoVAM and R3.

\end{abstract}

\section{Introduction}\label{sec:introduction}

\begin{figure*}[!t]
\centering
\includegraphics[width=\textwidth]{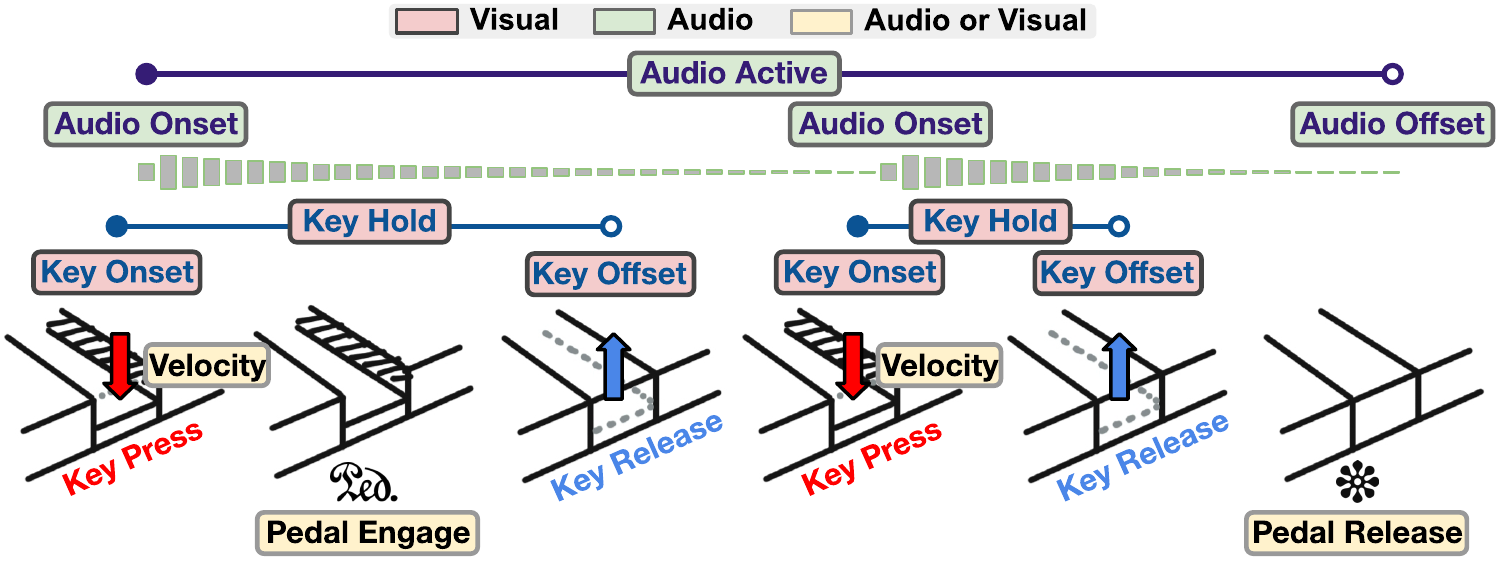}
\vspace{-20pt}
\caption{\textbf{Why video: physical key state decouples from audio under the sustain pedal.} Two strikes of the same key. \textcolor{red}{\textbf{Red~$\downarrow$}}: key press (onset, with velocity proportional to press force); \textcolor{blue}{\textbf{Blue~$\uparrow$}}: key release (offset). \emph{Onset} and \emph{velocity} are shared across modalities: a small hammer-travel delay separates \emph{key onset} from \emph{audio onset} but is absorbed by the evaluation tolerance. \emph{Key Hold} (video) tracks the physical state of the key, whereas \emph{Audio Active} (audio) extends through pedal sustain, so the two intervals agree only when the pedal is not engaged. Filled vs.\ open endpoints ($\bullet$\,/\,$\circ$) denote whether the boundary frame itself is included in the active interval (onset frame included, offset frame excluded).}
\label{fig:targets}
\vspace{-4pt}
\end{figure*}

Automatic Music Transcription (AMT) recovers symbolic note attributes such as pitch, onset, offset, and velocity from audio. Audio-based piano transcription has matured in estimating pitch, onset, and velocity~\cite{hawthorne2018, hawthorne2021, kong2021}, but \emph{offset} remains systematically confounded by the sustain pedal: sound persists after the key returns to rest, so audio systems conventionally extend note offsets to pedal release for both training and evaluation~\cite{hawthorne2018, kong2021, yan2021, yan2024}, with discrepancies from the MIDI \texttt{NoteOff} event that can reach several seconds.

Visual Piano Transcription (VPT) observes the keyboard directly: a pressed key is visually distinct from a released one regardless of pedal state, making the MIDI-encoded physical key state directly observable. VPT is also robust when audio is ambiguous, degraded, or absent, for example, in multi-instrument recordings, noisy or reverberant environments, and footage with missing or corrupted audio. Yet VPT remains \emph{underexplored}. Existing methods~\cite{koepke2020, su2020, zivanovic2025} process at most 0.2\,s of video context, leave offset accuracy substantially lower than onset accuracy, or do not report note-level velocity.

We present \textbf{V2N (Video to Notes)},\footnote{Code, trained checkpoints, and predicted MIDI:
\url{https://github.com/yonghyunk1m/V2N}.} the first complete VPT system: four task-specific heads (onset, offset, key hold, and velocity; Figure~\ref{fig:targets}) are trained jointly over 1\,s of video and supervised at every frame rather than only at the window center. Our contributions are:
\begin{itemize}
    \item \textbf{Video-only complete MIDI transcription.} V2N matches Li et al.~\cite{li2024avf} on PianoVAM Onset and surpasses all prior VPT on R3 Onset, substantially improves physical key-release prediction, and is the first video-only system to report note-level velocity F1, achieving state-of-the-art results on PianoVAM and R3.
    \item \textbf{Ablation-validated design choices.} (i)~\emph{Multi-frame loss} with a Conformer-style convolutional backbone~\cite{gulati2020} outperforms the single-frame Sight-to-Sound (S2S)~\cite{koepke2020} recipe on onset. (ii)~\emph{Multi-task heads} improve onset and offset accuracy; removing these heads (offset, key hold, velocity) reduces onset performance and collapses offset F1. (iii)~\emph{Offset-guided note decoding}, which terminates notes at offset-head peaks (with key hold as a fallback) rather than at a key hold (the video analogue of audio frame-activity) threshold alone, improves over the convention from audio-based piano transcription~\cite{hawthorne2018}.
\end{itemize}

We additionally analyze cross-dataset transfer and find that both V2N and prior VPT systems fail similarly, highlighting the limitations of fixed-geometry preprocessing.

\section{Related Work}\label{sec:related_work}

\subsection{Audio-based Piano Transcription}

Audio-based piano transcription is a mature field~\cite{benetos2019}. Onsets and Frames~\cite{hawthorne2018} established multi-task prediction of onsets, sustained audio activity, and velocity, whose complementarity dramatically improved transcription. Subsequent work refined the paradigm with Transformer architectures~\cite{hawthorne2021} and high-resolution onset/offset regression~\cite{kong2021} on MAESTRO~\cite{hawthorne2019}.
Neural semi-CRF event-based decoding~\cite{yan2021, yan2024} defines the current state of the art and, like its predecessors, extends note offsets through sustain-pedal intervals, so reported offset accuracy measures the end of sound rather than key release. We port this multi-task paradigm to vision, where onset cues differ (finger motion vs.\ spectral transients) and offset prediction becomes a direct observation rather than inferred from sound decay.

\subsection{Visual Piano Transcription}

Early VPT approaches detected pressed keys via background subtraction~\cite{suteparuk2014, akbari2015}. Sight-to-Sound (S2S)~\cite{koepke2020} introduced end-to-end learning with a ResNet-18~\cite{he2016} backbone on 5-frame windows. Su et al.'s Video2RollNet~\cite{su2020} (V2R) augmented the same ResNet-18 with multi-scale feature attention as the visual front-end of an audio-generation pipeline, predicting a binary pressed-key roll at the center of a 5-frame window. PPAN~\cite{zivanovic2025} applied a Vision Transformer~\cite{dosovitskiy2021} to 6-frame inputs of the R3 rehearsal dataset~\cite{cancinochacon2024}. These methods process at most 0.2\,s of temporal context, supervise only the center frame of each window, and leave offset accuracy well behind onset (PPAN: 45.9\% $+$Off on PianoVAM at 50\,ms, our re-evaluation). Velocity has been explored only with optical-flow CNNs~\cite{kang2019}, without note-level evaluation. The closest audio-visual counterpart, Li et al.~\cite{li2024avf}, fuses video with audio via cross-attention but derives offsets from acoustic frame activations and does not report note-level velocity.

\section{Methodology}\label{sec:proposed_method}

\subsection{Task Definition}\label{subsec:task_definition}

A piano note is defined by a key press (MIDI \texttt{NoteOn}) and a key release (\texttt{NoteOff}), together with pitch and velocity. Video captures \emph{mechanical} cues (keys pressing, being held, and returning to rest); audio captures \emph{acoustic} cues (hammer-strike transient and sound envelope).

With the sustain pedal engaged, sound persists long after key release, so audio-based piano transcription conventionally targets the end of the \emph{sound} rather than the \emph{key release}. Audio offsets can diverge from MIDI \texttt{NoteOff} by seconds.

We define four visual prediction targets grounded in key mechanics (Figure~\ref{fig:targets}); each is the head's continuous output, indexed by video frame $t \in \{1,\ldots,T\}$ and piano key $k \in \{1,\ldots,88\}$:
\begin{compactitem}
    \item \textbf{Onset} $o_{t,k} \in [0,1]$: key press initiation (shared with audio-based piano transcription).
    \item \textbf{Offset} $r_{t,k} \in [0,1]$: physical key release, aligned with MIDI \texttt{NoteOff}; audio-based piano transcription conventionally predicts the end of the sound (\emph{audio offset}), which coincides with key release only when the pedal is disengaged and otherwise extends to pedal release.
    \item \textbf{Key Hold} $f_{t,k} \in [0,1]$: key physically held over $[\text{onset}, \text{offset})$; the audio equivalent (\emph{active}) extends through pedal sustain.
    \item \textbf{Velocity} $v_{t,k} \in [0,1]$: per-key normalized MIDI velocity, regressed by a linear head with onset-masked loss.
\end{compactitem}
The probability heads $o, r, f$ are sigmoid-bounded and binarized at threshold $\tau{=}0.5$ during note decoding (Section~\ref{sec:experimental_setup}); the velocity head is linear (its raw output may exceed $[0,1]$) and is clipped to $[0,1]$ then rescaled to integers in $\{0,\ldots,127\}$ at inference (Section~\ref{subsec:loss}).

A \texttt{NoteOn} is then $(t^*, k^*, v_{t^*})$ at a peak in $o_{t,k}$, and \texttt{NoteOff} is the corresponding peak in $r_{t,k}$; the key hold signal densely labels the $[\text{onset}, \text{offset})$ span, and repeated same-key onsets yield independent note pairs while overlapping key hold labels merge via element-wise maximum.\footnote{We use ``frame'' exclusively for a video frame (40\,ms; all models in this paper operate at 25\,fps). This should not be confused with the ``frame'' terminology in audio-based piano transcription, which denotes per-step activity over a time-frequency representation (e.g., spectrogram or CQT).}

\subsection{Architecture}\label{subsec:architecture}

\begin{figure*}[!t]
\centering
\includegraphics[width=\textwidth]{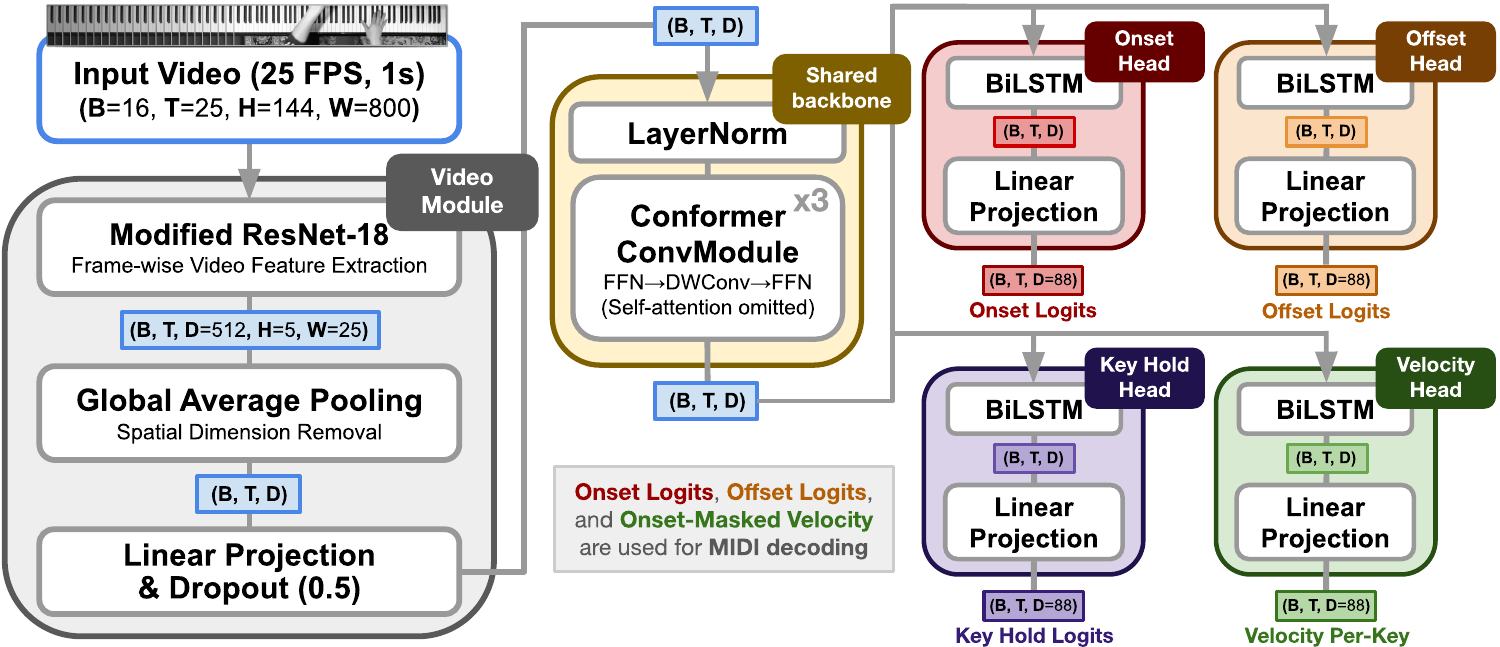}
\vspace{-14pt}
\caption{\textbf{V2N (Video to Notes) architecture.} Cropped, grayscale video frames pass through the video module: frame-wise S2S~\cite{koepke2020} feature extractor (ResNet-18~\cite{he2016} with a learned slope prior and 5-frame aggregation), followed by GAP and linear projection. A shared backbone of three Conformer ConvModule blocks (FFN$\to$DWConv$\to$FFN; self-attention omitted) feeds four parallel BiLSTM$\to$Linear heads for onset, offset, key hold, and velocity. At inference, onset peaks define note start (with per-onset velocity), offset peaks define note end; if key hold ends first, the note ends there instead.}
\label{fig:architecture}
\vspace{-4pt}
\end{figure*}

Our architecture (Figure~\ref{fig:architecture}) consists of three components: 
\begin{inparaenum}[(i)]
\item a visual feature extractor that converts each video frame into a per-frame feature vector, 
\item a shared temporal backbone that models dependencies across frames, and 
\item task-specific prediction heads for each output.
\end{inparaenum}

\textbf{Visual Feature Extractor.} We adopt the S2S feature extractor~\cite{koepke2020}: a ResNet-18~\cite{he2016} (grayscale, trained from scratch) augmented with a \emph{slope prior}, a learned 1D spatial encoding mapping each horizontal position to one of the 88 piano keys. For each frame, a 5-frame window (0.2\,s at 25\,fps) is processed jointly; global average pooling yields a $T{\times}512$ sequence, which a linear projection with dropout ($p{=}0.5$) maps to the backbone dimension.

\textbf{Temporal Backbone.} The projected features pass through LayerNorm followed by $L{=}3$ Conformer Conv\-Module blocks~\cite{gulati2020} (self-attention omitted): FFN $\rightarrow$ depthwise 1D convolution (kernel size 31) $\rightarrow$ FFN, with residual connections. Three stacked kernel-31 convolutions provide each output frame with a receptive field spanning all 25 frames of the 1\,s input without the quadratic cost of self-attention.

\textbf{Task-Specific Heads.} Four parallel heads each apply a bidirectional LSTM (hidden 256) followed by a linear projection. Onset, offset, and key hold heads output $T{\times}88$ logits, sigmoid-mapped to per-key per-frame probabilities; the velocity head regresses normalized MIDI velocity ($T{\times}88$) under a per-key mask at ground-truth onset positions. Key hold provides dense per-frame supervision over each $[\text{onset}, \text{offset})$ span, complementing the sparse onset/offset kernels and strengthening the backbone during training; at inference, offset peaks define note end; if key hold ends first, the note ends there instead (Table~\ref{tab:ablation_multitask}).

\textbf{Computational cost.} V2N uses 125\,GFLOPs and 28.2\,M parameters per 1\,s clip (S2S feature extractor accounts for 99\%; Conformer and BiLSTM heads add $<$1\%). With 0.5\,s overlapping stride, the amortized cost is ${\approx}250$\,GFLOPs per second of input video, or 40\,ms of GPU time on an NVIDIA RTX 5080 (bfloat16; RTF~${\approx}0.04$). This undercuts Li et al.~\cite{li2024avf}'s video branch (${\sim}430$\,GFLOPs/s, ${\sim}1.7{\times}$) and is comparable in wall-clock to PPAN~\cite{zivanovic2025} (9.7\,GFLOPs per 6-frame call) despite higher FLOPs, so the 1\,s window incurs no practical wall-clock penalty.

\subsection{Training}\label{subsec:loss}

Onset and offset labels are soft triangular kernels centered at each event, linearly decaying from 1.0 at the event frame to 0 at $\pm\ell$ frames ($\ell{=}2$); this tolerates small labeling jitter and spreads the gradient signal over neighboring frames instead of a single positive surrounded by negatives. Key hold labels are binary 1 throughout $[\text{onset}, \text{offset})$. MIDI velocity is normalized to $[0,1]$ for training (dividing by 127) and at inference is rescaled by 127, clipped to $[0,127]$, and rounded to integers in $\{0,\ldots,127\}$ for MIDI export.
\begin{equation}
    L_\text{vel} = \frac{\sum_{t,k} m_{t,k} (v_{t,k} - v^*_{t,k})^2}{\sum_{t,k} m_{t,k}}
\end{equation}
where $m_{t,k} \in \{0,1\}$ is a per-key onset mask (1 at ground-truth onset positions, 0 elsewhere). Task weights are onset$=$2, others$=$1, emphasizing the onset head as in Onsets-and-Frames~\cite{hawthorne2018}:
\begin{equation}
    L_\text{total} = 2 L_\text{onset} + L_\text{off} + L_\text{khold} + L_\text{vel}
\end{equation}

At training time, 1\,s segments are drawn densely at a 1-frame stride and shuffled across videos, so every frame index appears as the center of some segment and at every other relative offset across epochs. At inference we process the full recording in 1\,s segments with 0.5\,s stride (50\% overlap), retaining only the central 0.5\,s of predictions from each segment to avoid boundary artifacts where the receptive field is truncated at segment edges.

\begin{table*}[!t]
\centering
\small
\setlength{\tabcolsep}{7pt}
\begin{tabular}{@{}l c cc cc cc cc@{}}
  \toprule
  & \textbf{Frame} & \multicolumn{2}{c}{\textbf{Onset}} & \multicolumn{2}{c}{\textbf{$+$Off}} & \multicolumn{2}{c}{\textbf{$+$Vel}} & \multicolumn{2}{c}{\textbf{$+$Off$+$Vel}} \\
  \cmidrule(lr){3-4} \cmidrule(lr){5-6} \cmidrule(lr){7-8} \cmidrule(lr){9-10}
  \textbf{Model} & (60\,Hz) & \textbf{50\,ms} & \textbf{100\,ms} & \textbf{50\,ms} & \textbf{100\,ms} & \textbf{50\,ms} & \textbf{100\,ms} & \textbf{50\,ms} & \textbf{100\,ms} \\
  \midrule
  \multicolumn{10}{l}{\textit{Trained on PianoVAM $\rightarrow$ Test: PianoVAM}} \\
  S2S$^\dagger$~\cite{koepke2020} & 41.7 & 60.1 & 94.5 & 21.8 & 49.6 & (39.4) & (63.2) & (14.3) & (32.7) \\
  V2R$^\dagger$~\cite{su2020} & 82.2 & 86.5 & 89.8 & 55.4 & 81.8 & (58.5) & (60.4) & (36.8) & (54.6) \\
  Li et al.$^\dagger$~\cite{li2024avf} & (25.2) & 94.2 & \textbf{97.4} & (21.0) & (40.0) & (63.6) & (65.2) & (12.8) & (25.8) \\
  PPAN$^\dagger$~\cite{zivanovic2025} & 80.4 & 84.9 & 94.1 & 45.9 & 82.8 & (57.0) & (62.5) & (29.7) & (55.3) \\
  V2N (Ours) & \textbf{90.9}\rlap{$^{*}$} & \textbf{94.7} & 97.0 & \textbf{89.5}\rlap{$^{*}$} & \textbf{95.5}\rlap{$^{*}$} & \textbf{82.8}\rlap{$^{*}$} & \textbf{84.5}\rlap{$^{*}$} & \textbf{78.3}\rlap{$^{*}$} & \textbf{83.2}\rlap{$^{*}$} \\
  \midrule
  \multicolumn{10}{l}{\textit{Trained on R3s+R3x $\rightarrow$ Test: R3s}} \\
  S2S$^\dagger$~\cite{koepke2020} & 60.5 & 44.0 & 84.7 & 17.2 & 64.7 & (24.9) & (46.5) & (9.8) & (35.2) \\
  V2R$^\dagger$~\cite{su2020} & 54.7 & 38.7 & 76.1 & 15.8 & 54.9 & (21.6) & (41.4) & (8.8) & (29.6) \\
  PPAN$^\dagger$~\cite{zivanovic2025} & 57.0 & 40.7 & 83.4 & 14.9 & 60.3 & (23.4) & (46.1) & (8.7) & (33.2) \\
  V2N (Ours) & \textbf{80.4}\rlap{$^{*}$} & \textbf{78.8}\rlap{$^{*}$} & \textbf{91.7}\rlap{$^{*}$} & \textbf{69.5}\rlap{$^{*}$} & \textbf{88.9}\rlap{$^{*}$} & \textbf{57.9}\rlap{$^{*}$} & \textbf{65.9}\rlap{$^{*}$} & \textbf{51.2}\rlap{$^{*}$} & \textbf{64.1}\rlap{$^{*}$} \\
  \midrule
  \multicolumn{10}{l}{\textit{Trained on R3s+R3x $\rightarrow$ Test: R3x}} \\
  S2S$^\dagger$~\cite{koepke2020} & 58.7 & 46.6 & 83.9 & 14.7 & 63.6 & (33.5) & (57.1) & (10.3) & (44.2) \\
  V2R$^\dagger$~\cite{su2020} & 57.0 & 46.4 & 80.8 & 14.9 & 60.8 & (33.7) & (55.7) & (10.3) & (42.6) \\
  PPAN$^\dagger$~\cite{zivanovic2025} & 56.6 & 47.7 & 83.2 & 13.9 & 59.3 & (33.7) & (57.0) & (9.3) & (41.3) \\
  V2N (Ours) & \textbf{76.5}\rlap{$^{*}$} & \textbf{73.3}\rlap{$^{*}$} & \textbf{86.6}\rlap{$^{*}$} & \textbf{68.5}\rlap{$^{*}$} & \textbf{85.2}\rlap{$^{*}$} & \textbf{59.7}\rlap{$^{*}$} & \textbf{69.8}\rlap{$^{*}$} & \textbf{56.6}\rlap{$^{*}$} & \textbf{69.0}\rlap{$^{*}$} \\
  \bottomrule
\end{tabular}
\vspace{-2pt}
\caption{\textbf{Main results} (F1 \%; Frame F1 on a 60\,Hz grid; onset and offset matched within the stated $\tau \in \{50, 100\}$\,ms under our strict convention; definitions in Section~\ref{sec:experimental_setup}). \textbf{Bold} = best per column within each test split. $^\dagger$: trained from scratch (S2S and V2R via the reproductions in~\cite{zivanovic2025}; Li et al.\ video branch only and PPAN with authors' own code). Parenthesized values indicate a degenerate/absent baseline head: constant velocity for all baselines, and near-zero-duration offsets for Li et al. $^{*}$: $p{<}0.01$ vs.\ runner-up, paired Wilcoxon one-sided ($n{=}9$ PianoVAM, $74$ R3s, $107$ R3x).}
\label{tab:main}
\vspace{-4pt}
\end{table*}

\section{Experimental Setup}\label{sec:experimental_setup}

\subsection{Datasets}

\textbf{PianoVAM}~\cite{pianovam2025} contains 107 top-view piano videos with synchronized MIDI from a Yamaha Disklavier, recorded during amateur practice sessions under naturalistic conditions. We follow the proposed splits in PianoVAM v1.1's \texttt{metadata.json}: \texttt{train}$+$\texttt{ext-train} (81 recordings) for training, \texttt{valid} (9) for validation, and \texttt{test} (9 recordings, 1.63\,h, 42{,}241 notes) for evaluation; we exclude 8 recordings that the release assigns to \texttt{special(blurry)} and \texttt{special(4hands)} splits. Each video is preprocessed with a perspective transform using annotated keyboard corners, producing $800{\times}144$ grayscale frames at 25\,fps. The frame height includes the keyboard and a bottom margin capturing hand and wrist motion for velocity cues.

\textbf{R3}~\cite{cancinochacon2024, zivanovic2025} consists of 31\,h of practice recordings from 62 sessions by two professional pianists, captured with GoPro cameras (primarily 60\,fps, downsampled to 25\,fps). It is divided into R3s (Rachmaninoff Piano Concerto No.~3; 219 train / 74 test) and R3x (various Western classical repertoire; 495 train / 107 test). Videos are preprocessed with a perspective transform using bounding-box annotations and rotation correction to the same $800{\times}144$ format. R3 is more challenging than PianoVAM: it spans multiple pianos, camera positions, lighting, and advanced repertoire.

\subsection{Training Configuration}

\textbf{V2N.} Trained with AdamW~\cite{loshchilov2019} ($\text{lr}{=}5{\times}10^{-4}$, weight decay $0.01$) and a cosine schedule with 5\% linear warmup, in bfloat16 with an effective batch size of 16 (via gradient accumulation). To fit the 1\,s window on a single GPU, we apply \emph{gradient sampling}: backpropagation flows through a random 50\% of the 25 input frames while the remainder are forward-passed under \texttt{torch.no\_grad}, roughly halving backward-pass memory without reducing frame-level supervision coverage. PianoVAM models use 100k optimizer steps and R3 models use 200k steps (convergence was slower on the more heterogeneous R3 data). All reported numbers come from the final-step checkpoint. For augmentation, we adopt a subset of the PPAN training recipe~\cite{zivanovic2025}: brightness jitter ($\pm$10\%, $p{=}0.4$), random rotation ($\pm$0.2$^\circ$, $p{=}0.4$), random erasing ($p{=}0.5$), and Gaussian noise ($\sigma{=}0.1$, $p{=}0.4$).

\textbf{Baselines.} S2S~\cite{koepke2020}, V2R~\cite{su2020}, Li et al.~\cite{li2024avf}, and PPAN~\cite{zivanovic2025} are retrained with each author's default hyperparameters and augmentations. Li et al.\ and PPAN use official code. S2S and V2R use the CNN and CNN-V2R reproductions from the PPAN codebase (5-frame window, image height$\times$width $160{\times}800$ and $138{\times}776$, respectively). For Li et al., we train and evaluate only the video branch (no audio input), so all baselines are video-only. On R3, S2S, V2R, and PPAN share the PPAN recipe (AdamW, 10 epochs on R3s+R3x, PPAN-default augmentations).

\begin{table*}[!tbp]
\centering
\small
\setlength{\tabcolsep}{6pt}
\begin{tabular*}{\textwidth}{@{\extracolsep{\fill}}l l c cc cc cc cc@{}}
  \toprule
  \multirow{2}{*}{\textbf{Training}} & \multirow{2}{*}{\textbf{Inference}} & \textbf{Frame} & \multicolumn{2}{c}{\textbf{Onset}} & \multicolumn{2}{c}{\textbf{$+$Off}} & \multicolumn{2}{c}{\textbf{$+$Vel}} & \multicolumn{2}{c}{\textbf{$+$Off$+$Vel}} \\
  \cmidrule(lr){4-5} \cmidrule(lr){6-7} \cmidrule(lr){8-9} \cmidrule(lr){10-11}
  & & (60\,Hz) & 50\,ms & 100\,ms & 50\,ms & 100\,ms & 50\,ms & 100\,ms & 50\,ms & 100\,ms \\
  \midrule
  Kh           & Kh            & 90.1 & 92.8 & 95.2 & 81.4 & 93.1 & (62.3) & (63.8) & (55.5) & (62.4) \\
  Kh, V        & Kh, V         & 90.6 & 92.7 & 96.1 & 85.7 & 94.1 & 80.6 & 82.8 & 74.6 & 81.0 \\
  Kh, V, On    & Kh, V         & 90.2 & 93.3 & \underline{96.8} & 85.9 & 95.0 & 82.1 & \underline{84.9} & 75.8 & \textbf{83.4} \\
  Kh, V, On    & Kh, V, On     & 90.4 & \underline{94.3} & \textbf{97.0} & 86.7 & 95.2 & \textbf{82.8} & \textbf{85.0} & 76.3 & \textbf{83.4} \\
  Kh, V, On, Off & Kh, V, On   & \underline{90.7} & \textbf{94.7} & \textbf{97.0} & \underline{87.8} & \underline{95.4} & \underline{82.5} & 84.3 & \underline{76.7} & 83.0 \\
  \textbf{Kh, V, On, Off} & \textbf{Kh, V, On, Off} & \textbf{90.9} & \textbf{94.7} & \textbf{97.0} & \textbf{89.5} & \textbf{95.5} & \textbf{82.8} & 84.5 & \textbf{78.3} & \underline{83.2} \\
  \bottomrule
\end{tabular*}
\vspace{-4pt}
\caption{\textbf{Task-head and decoder ablation on PianoVAM} (F1 \%; 1\,s window; same $\tau$ convention as Table~\ref{tab:main}). Training lists heads jointly trained; Inference lists heads used at MIDI decoding (note end is the earlier of an offset peak or a key hold drop, when both are available). Heads: Kh = key hold, V = velocity, On = onset, Off = offset. The bold bottom row is our final V2N configuration. \textbf{Bold} values = best per column, \underline{underline} = 2nd-best per column.}
\label{tab:ablation_multitask}
\vspace{-4pt}
\end{table*}

\begin{table}[!tbp]
\centering
\small
\setlength{\tabcolsep}{5pt}
\begin{tabular*}{\linewidth}{@{\extracolsep{\fill}}l c ccc@{}}
  \toprule
  \textbf{Variant} & \textbf{Window} & \textbf{Onset} & \textbf{$+$Off} & \textbf{$+$Vel} \\
  \midrule
  Ours (center-frame loss) & 0.2\,s & 90.5 & 78.5 & 77.4 \\
  $+$ multi-frame loss     & 0.2\,s & 91.8 & 83.0 & 78.5 \\
  $+$ sequence model       & 0.2\,s & 94.3 & 88.7 & 80.5 \\
  $+$ longer context (V2N) & 1.0\,s & \textbf{94.7} & \textbf{89.5} & \textbf{82.8} \\
  \bottomrule
\end{tabular*}
\vspace{-6pt}
\caption{\textbf{Architecture ablation on PianoVAM} (F1 \%; onset and offset matched within $\tau{=}50$\,ms, as in Table~\ref{tab:main}). Each row adds one component over the previous.}
\label{tab:ablation_pv}
\vspace{-4pt}
\end{table}

\subsection{Evaluation Protocol}

We decode frame-level predictions into MIDI note events using the onset, offset, key hold, and velocity heads. A \emph{peak} is a rising-edge transition of the binarized probability on a given key (the first above-threshold frame after an off frame). Each note begins at an onset peak above $\tau_\text{on}{=}0.5$ and is extended forward until the next offset peak above $\tau_\text{off}{=}0.5$ on the same key, or until both onset and key hold probabilities fall below threshold, whichever comes first. Predicted MIDI is evaluated against ground truth with \texttt{mir\_eval}~\cite{raffel2014}. We report one frame-level and four additive note-level F1 metrics (\texttt{mir\_eval} function names in parentheses):
\begin{itemize}
\setlength{\itemsep}{1pt}
\setlength{\parsep}{0pt}
    \item \textbf{Frame} ({\small\texttt{multipitch}}): per-frame multipitch F1 after rasterizing predicted and ground-truth notes on a 60\,Hz grid, matching the primary original video frame rate. Independent of onset tolerance.
    \item \textbf{Onset} ({\small\texttt{note}}): pitch and onset match within an onset tolerance $\tau$ (50\,ms or 100\,ms).
    \item \textbf{$+$Off} ({\small\texttt{note\_with\_offsets}}): offset match within the \emph{same} $\tau$ as onset rather than {\small\texttt{mir\_eval}}'s $\max(0.2d, 50\,\mathrm{ms})$, reflecting key release.
    \item \textbf{$+$Vel} ({\small\texttt{note\_with\_velocity}}): Onset match plus velocity match within 10\% ({\small\texttt{mir\_eval}} rescales velocity to its L2-optimal global scalar). \emph{Not cumulative with $+$Off}.
    \item \textbf{$+$Off$+$Vel} ({\small\texttt{note\_with\_offsets\_and\_velocity}}): the strictest metric; all four MIDI attributes (pitch, onset, offset, velocity) must match.
\end{itemize}
We report note-level results at both 50\,ms (standard in audio-based piano transcription) and 100\,ms (matching the protocol of~\cite{zivanovic2025}) onset tolerances; under the offset convention above, the offset tolerance tracks the onset tolerance. Predicted note timestamps are quantized to the 25\,fps video grid (40\,ms resolution), while ground-truth MIDI timestamps are continuous; the 50\,ms onset/offset tolerance absorbs up to 40\,ms of quantization error, so evaluation is not bottlenecked by the video frame rate. The sustain pedal (MIDI CC64) is not predicted by V2N or any baseline.

\begin{figure*}[!tbp]
\centering
\includegraphics[width=\textwidth]{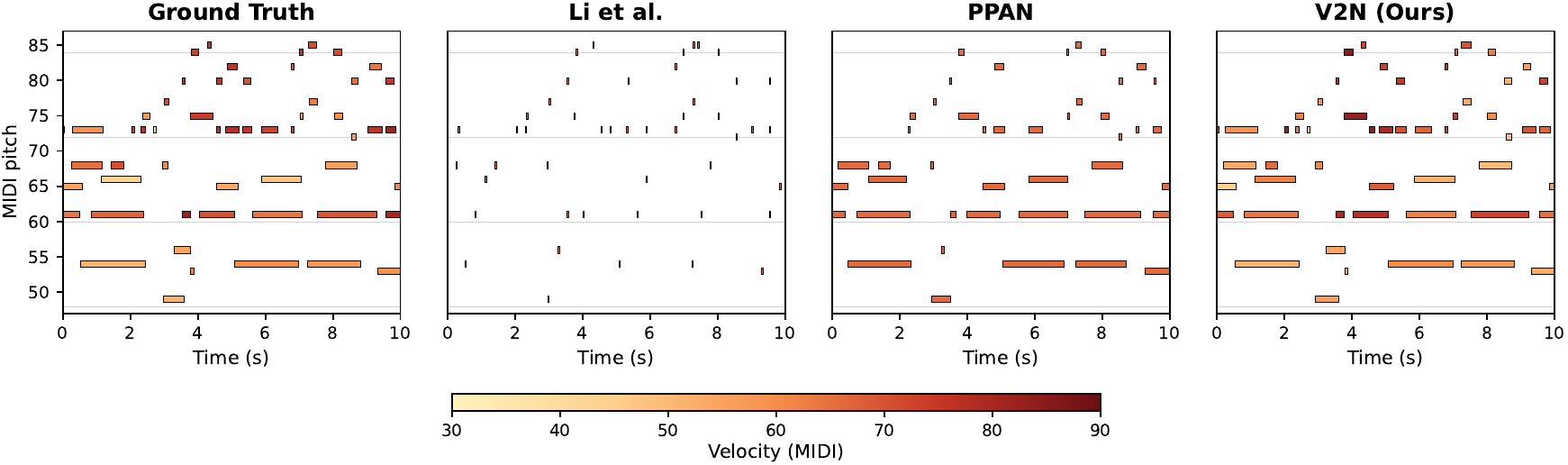}
\vspace{-22pt}
\caption{\textbf{Qualitative piano-roll comparison.} 10\,s PianoVAM excerpt (Yiruma, \textit{Kiss the Rain}; test recording \texttt{2024-02-17\_21-44-37}, t$=$130--140\,s). Rectangle: pitch (height) $\times$ note duration (width), filled by MIDI velocity. Li et al.\ and PPAN have no velocity head; their notes are rendered at the segment's ground-truth mean velocity, matching \texttt{mir\_eval}'s L2-optimal rescaling for constant-velocity predictions. V2N reconstructs both durations and dynamics.}
\label{fig:pianoroll}
\vspace{-4pt}
\end{figure*}

\section{Results}\label{sec:results}

\subsection{Main Results}

Table~\ref{tab:main} compares V2N against the baselines defined in Section~\ref{sec:experimental_setup}. V2N is our default full-configuration model (onset, offset, key hold, velocity heads).

On PianoVAM, V2N matches Li et al.\ on Onset (within 0.5\,\%p at both tolerances) and surpasses all baselines on every offset-dependent metric. The largest gains appear on the physical key-release metrics ($+$Off and $+$Off$+$Vel): dedicated offset supervision with 1\,s of context nearly doubles $+$Off F1 (PPAN: 45.9 $\to$ V2N: 89.5 at 50\,ms) and yields $+$48.6\,\%p on $+$Off$+$Vel (29.7 $\to$ 78.3) over PPAN, and V2N is the only system achieving high accuracy across all four MIDI attributes (pitch, onset, offset, velocity).

On R3, V2N surpasses prior VPT on Onset at both splits and tolerances, and the gap on offset-dependent metrics grows further: onset-centric baselines fail on $+$Off despite competitive onset accuracy, confirming that physical key release requires dedicated supervision rather than arising as a by-product of key hold (frame-activity) modeling.

Our retrained S2S and V2R baselines reproduce the published R3s/R3x Onset numbers within roughly one percentage point, so the gap V2N opens on offset-dependent metrics reflects model design, not training differences.

Figure~\ref{fig:pianoroll} illustrates the qualitative consequence of these performance differences on a representative PianoVAM excerpt: Li et al.'s offsets collapse to near-zero duration, PPAN produces plausible note spans but a constant MIDI velocity, whereas V2N reconstructs both the note durations and the dynamic contour of the ground truth.

\subsection{Ablation Studies}

We organize the ablations to mirror the title: first \emph{multi-task} heads (Table~\ref{tab:ablation_multitask}), then \emph{multi-frame} context (Table~\ref{tab:ablation_pv}).

\smallskip\noindent\textbf{Why multi-task heads?} Table~\ref{tab:ablation_multitask} shows that additional heads help both as training supervision and as decoding signals. Adding velocity supervision and decoding raises $+$Vel F1 from the constant-velocity proxy to 80.6 at 50\,ms. Adding onset as an auxiliary target improves most note metrics before the onset head is used (row 3 vs.\ row 2), and using it at decoding further improves Onset F1 (93.3 $\to$ 94.3; row 4 vs.\ row 3). Finally, adding offset supervision improves offset-dependent metrics before the offset head is used (row 5 vs.\ row 4), and offset-guided decoding yields the best $+$Off and $+$Off$+$Vel scores (row 6). Thus, the heads provide complementary supervision and inference-time cues for complete MIDI prediction.

\smallskip\noindent\textbf{Why multi-frame loss?} Our starting variant (Table~\ref{tab:ablation_pv} row 1) follows the S2S~\cite{koepke2020} recipe of supervising only the \emph{center} frame of each 5-frame window. Extending the loss to all five frames (row 2) yields $+$1.3\,\%p Onset and $+$4.5\,\%p $+$Off, since the model now sees ${\sim}5{\times}$ more supervised signal per segment at zero inference cost.

\smallskip\noindent\textbf{Why sequence modeling?} Adding a Conformer ConvMo\-dule backbone over the 5-frame window (Table~\ref{tab:ablation_pv} row 3) is the largest single gain: $+$2.5\,\%p Onset and $+$5.7\,\%p $+$Off. A purely spatial S2S head cannot distinguish \emph{pressed} from \emph{held} keys; sequence modeling resolves this ambiguity.

\smallskip\noindent\textbf{Why a 1\,s input window?} Extending the window from 0.2\,s to 1.0\,s (Table~\ref{tab:ablation_pv} row 4) adds only $+$0.4/$+$0.8\,\%p (Onset/$+$Off) on PianoVAM but is decisive on R3: V2N reaches 91.7\% R3s Onset F1 at 100\,ms (Table~\ref{tab:main}) and enables offset prediction (+Off 88.9\%) that shorter-window, onset-only baselines cannot provide. R3's advanced repertoire and heterogeneous conditions demand longer context, so we use 1.0\,s as the default on both datasets.

\smallskip\noindent\textbf{Why parallel heads?} Cascaded variants from audio-transcription designs~\cite{hawthorne2018,kong2021} yielded no meaningful improvement, so V2N uses the simpler parallel-head design.

\subsection{Cross-Dataset Transfer}\label{subsec:cross_dataset}

\begin{table}[!tbp]
\centering
\small
\setlength{\tabcolsep}{5pt}
\begin{tabular*}{\linewidth}{@{\extracolsep{\fill}}ll cccc@{}}
  \toprule
  \textbf{Train} & \textbf{Test} & \textbf{S2S} & \textbf{V2R} & \textbf{PPAN} & \textbf{V2N} \\
  \midrule
  PianoVAM & R3s      & 0.6/2.4   & 0.3/0.5   & 0.0/0.0   & 1.5/5.6 \\
  PianoVAM & R3x      & 0.2/1.8   & 0.0/0.0   & 0.0/0.0   & 0.9/2.7 \\
  R3s$+$R3x & PianoVAM & 0.1/0.1  & 0.0/0.0  & 0.3/0.6 & 1.1/4.9 \\
  \bottomrule
\end{tabular*}
\vspace{-6pt}
\caption{\textbf{Cross-dataset transfer} (Onset F1 \%; 50\,ms / 100\,ms). Transfer between PianoVAM and R3 collapses everywhere, reflecting shared fixed-geometry preprocessing.}
\label{tab:cross}
\vspace{-4pt}
\end{table}

Table~\ref{tab:cross} shows that Onset F1 under dataset swaps collapses to near zero for every system. The root cause is geometric: the $800{\times}144$ perspective warp standardizes the \emph{pixel grid}, not the \emph{keyboard within it}; median keyboard width is ${\sim}784$\,px on R3 vs.\ ${\sim}606$\,px on PianoVAM, so the same column indexes a different key across datasets. Cross-dataset VPT likely requires geometry-invariant representations, as also observed in audio-based AMT~\cite{martak2024}. We leave robustness to unseen camera geometries to future work.

\section{Discussion}\label{sec:discussion}

\smallskip\noindent\textbf{R3 synchronization issues.} Cross-correlating MIDI onset trains with video audio energy (10\,ms resolution, $\pm$3\,s search) reveals systematic video--MIDI offsets exceeding 200\,ms in 70 of 895 R3 files, all with MIDI preceding video. Since this affects all models equally, we report results on the original test split. Excluding the 10 sync-affected R3x test files raises V2N F1 at 100\,ms by $+7.7$\,\%p Onset (86.6 $\to$ 94.3), $+7.9$\,\%p $+$Off (85.2 $\to$ 93.2), $+6.6$\,\%p $+$Vel (69.8 $\to$ 76.4), and $+6.7$\,\%p $+$Off$+$Vel (69.0 $\to$ 75.7). Per-file sync offsets are released with the code (Section~\ref{sec:introduction}).

\smallskip\noindent\textbf{Limitations.} The current work has two main limitations:
\begin{inparaenum}[(i)]
\item cross-dataset transfer fails almost completely: the learned pixel-to-key mapping does not generalize across camera geometries, even after perspective-transform normalization, and
\item V2N does not estimate sustain-pedal control changes.
\end{inparaenum}
Future work should address geometry-invariant keyboard localization, joint sync alignment for noisy training data, sustain-pedal estimation, and audio-visual fusion building on V2N's visual contributions.

\section{Conclusion}\label{sec:conclusion}

We presented V2N, a visual piano transcription system that jointly predicts onsets, offsets, key hold states, and velocity through multi-task, multi-frame modeling. V2N substantially improves physical key-release accuracy over prior video baselines, whose offsets either collapse to near-zero duration or are inferred post-hoc from key hold activations, and is the first VPT system to report note-level velocity. Ablations identify three design choices behind complete note-level transcription from video: a dedicated offset head, multi-task training, and offset-guided decoding. Cross-dataset transfer between PianoVAM and R3 collapses in every direction, indicating that current VPT models are tightly coupled to their training preprocessing; geometry-invariant representations and dataset-agnostic keyboard localization remain open problems.

\section{Acknowledgements}
This work was supported by the National Research Foundation of Korea (NRF) grant funded by the Korea government (MSIT) under Grant RS-2023-NR077289. 

\bibliography{references}

\end{document}